# Fabrication of the DESI Corrector Lenses

Timothy N. Miller*[a], Robert W. Besuner[a], Michael E. Levi[a], Michael Lampton[a], Patrick Jelinsky[a], Henry Heetderks[a], David J. Schlegel[a], Jerry Edelstein[a], Peter Doel[b], David Brooks[b], Stephen Kent[c], Gary Poczulp[d], Michael J. Sholl[†a]

[a] Lawrence Berkeley National Laboratory, 1 Cyclotron Rd, Berkeley, CA USA, 94720;
[b] University College London, Gower St, London WC1E 6BT, United Kingdom;
[c] Fermilab/MS127, P.O. Box 500, Batavia, IL USA, 60501;
[d] National Optical Astronomy Observatory, 950 N Cherry Ave, Tucson, AZ USA, 85719

## ABSTRACT

The Dark Energy Spectroscopic Instrument (DESI) is under construction to measure the expansion history of the Universe using the Baryon Acoustic Oscillation technique. The spectra of 35 million galaxies and quasars over 14000 square degrees will be measured during the life of the experiment. A new prime focus corrector for the KPNO Mayall telescope will deliver light to 5000 fiber optic positioners. The fibers in turn feed ten broad-band spectrographs.

We describe the DESI corrector optics, a series of six fused silica and borosilicate lenses. The lens diameters range from 0.8 to 1.1 meters, and their weights 84 to 237 kg. Most lens surfaces are spherical, and two are challenging 10$^{th}$-order polynomial aspheres. The lenses have been successfully polished and treated with an antireflection coating at multiple subcontractors, and are now being integrated into the DESI corrector barrel assembly at University College London.

We describe the final performance of the lenses in terms of their various parameters, including surface figure, homogeneity, and others, and compare their final performance against the demanding DESI corrector requirements. Also we describe the reoptimization of the lens spacing in their corrector barrel after their final measurements are known. Finally we assess the performance of the corrector as a whole, compared to early budgeted estimates.

## 1. INTRODUCTION

The Dark Energy Spectroscopic Instrument (DESI) will be used to perform an optical/near-infrared survey of 35 million galaxies in order to answer astronomical questions about the nature of dark energy in the universe.[1] DESI will be installed at the Kitt Peak National Observatory on the 4-meter Mayall telescope starting in 2018. The DESI instrument consists of a new corrector assembly at the focus of the Mayall primary mirror, a fiber run at the telescope image surface that carries the light away from the telescope, and an array of spectrometers that performs spectroscopy on the astronomical light.

The existing Mayall corrector will be replaced in 2018 with the new DESI corrector that will allow excellent imaging of the sky to a focal plane over a 3.2-degree diameter field of view.[2] This corrector requires a group of meter-class lenses for its operation.

The six corrector lenses have been fully fabricated and coated, and are being integrated into the corrector assembly. This talk gives an overview of the DESI corrector lens design, along with a brief status of the lenses. We then describe the parameters of the final fabricated lenses as compared to their specifications, and discuss the verification methods for the parameters. Finally we describe the respacing of the lenses in the design based on their final measured parameters, and assess the imaging performance of the complete lens assembly.

* timmiller@lbl.gov ; 510-486-6340

## 2. OVERVIEW OF THE DESI CORRECTOR

Figure 1 shows the overall DESI instrument on the Mayall telescope, with the new DESI corrector at the Mayall prime focus. Figure 2 shows a cutaway view of the DESI corrector. The corrector contains six large lenses that are mounted in athermal cells and positioned precisely in a stable barrel.[3] Figure 3 shows the layout of the DESI corrector optical design, a prime-focus corrector design that reimages the sky onto a curved focal surface, where the DESI fiber array carries light to the spectrographs.[4]

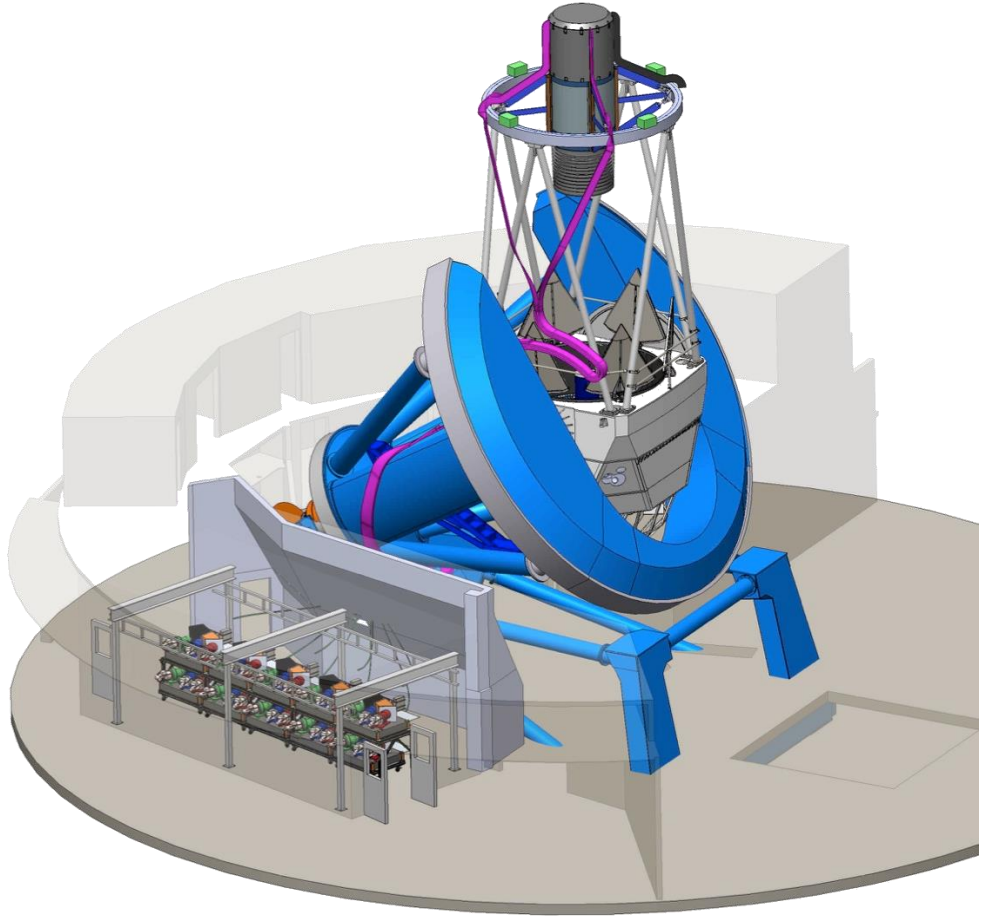

Figure 1. The Mayall telescope and the DESI instrument.

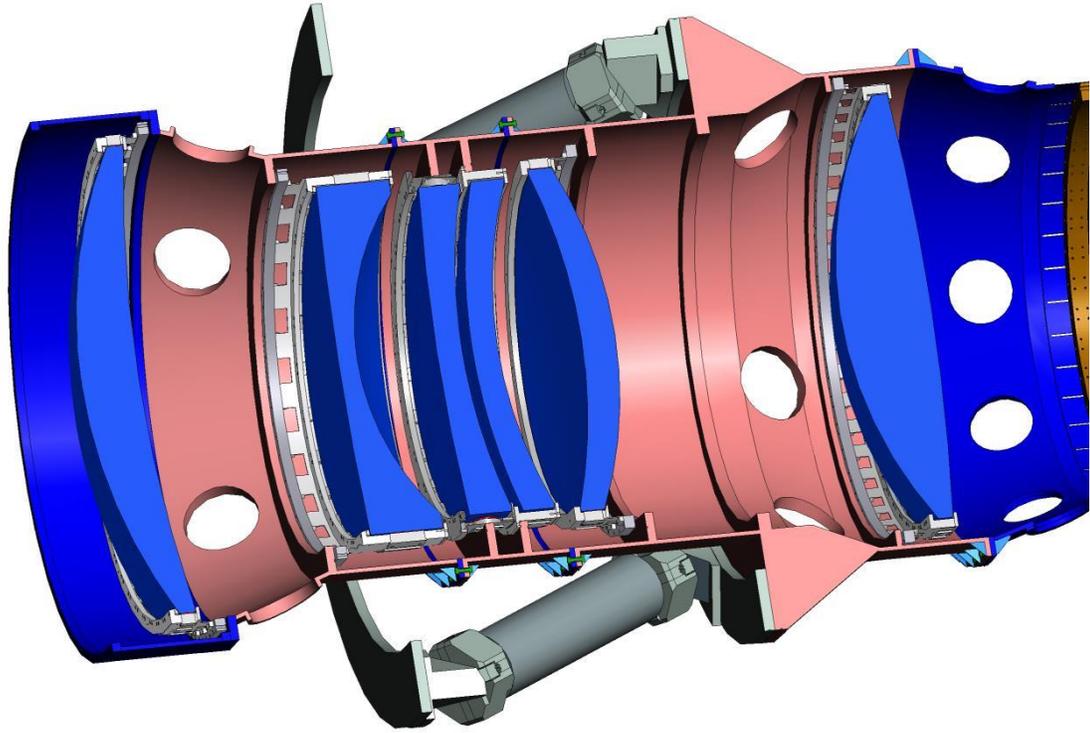

Figure 2. The DESI Corrector assembly, in cutaway view. The six lenses are shown in blue.

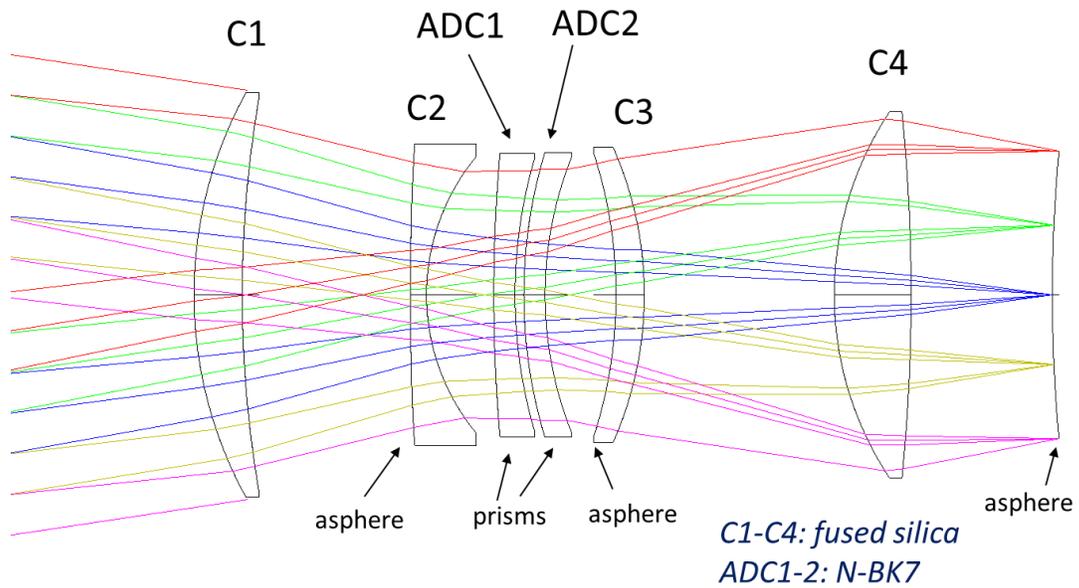

Figure 3. The DESI Corrector optical design.

The optical design achieves excellent imaging performance over a large 3.2-degree field of view and wide 360-980 nm bandpass. Furthermore the corrector constrains the angle of incidence at the focal surface to be close to normal, in order to maximize light captured by the fiber inputs there. It also includes an atmospheric dispersion corrector (ADC), two slightly wedged lenses that can be independently rotated to counteract the effect of wavelength-dependent dispersion by the atmosphere up to Zenith angles of 60 degrees. Finally the design minimized ghost reflections between surfaces.

Table 1 lists some parameters of the six corrector lenses. These lenses are demanding, in that they are large (up to 1.1m diameter), massive (up to 237 kg), and have rigorous specifications for such parameters as inhomogeneity and surface figure quality. Furthermore two of the lens surfaces are 10$^{th}$-order aspheres. However, the design was developed and optimized with an eye on reasonable fabrication. The materials used are only fused silica and borosilicate, two common glasses, chosen for their relative ease of availability and excellent internal quality. Vendors were contacted during the design process for their input about what design parameters and tolerances were within their capabilities.

Table 1. Main parameters of the DESI corrector lenses.

| Lens | Diameter (mm) | Mass (kg) | Center thickness (mm) | Material |
|---|---|---|---|---|
| C1 | 1140 | 201 | 136.4 | Fused Silica |
| C2 | 850 | 151 | 45 | Fused Silica |
| ADC1 | 800 | 102 | 60 | N-BK7 |
| ADC2 | 804 | 89 | 60 | S-BSL7 |
| C3 | 834 | 84 | 80 | Fused Silica |
| C4 | 1034 | 237 | 217 | Fused Silica |

## 3. CORRECTOR STATUS

Despite their demanding parameters, the lenses are still within the capability of modern optical fabrication shops. Progress on the fabrication of the lenses has been previously reported[5], and since then all six have been successfully polished and accepted by DESI.

Custom anti-reflection coatings were developed by Viavi Solutions, and successfully applied to all twelve lens surfaces; for details refer to the talk in this conference[6]. The lenses are currently at University College London, where they are being integrated into individual athermal cells, and the cells aligned into the corrector barrel[7].

In mid-2018 the corrector will be shipped from London to Kitt Peak for integration into the Mayall telescope, with early commissioning starting in fall of 2018.

## 4. LENS PERFORMANCE VS. SPECIFICATION

All polished lenses went through final testing after the polishing was complete to verify whether they met every specification within tolerance. Table 2 lists the primary specifications for each lens, and the final values of that specification for the completed lenses, verified by final measurements. The lenses meet almost all of their requirements by a significant margin. An exception is the rms slope error for the C2A apshere: DESI accepted a noncompliance on this particular spec, after determining by modeling that the performance hit due to the estimated error was acceptably small. Otherwise all specs are met by the blank and polishing vendors. Note that the vendor's method of measuring slope error in the case of ADC1 and ADC2 (a structure function) led to a curve rather than a single value; in all cases the curve met spec.

Table 2. The primary specifications for each lens, and the final values for lenses that have been completed.

| Lens | Substrate Inhomogeneity (ppm) * | | Radius tolerance (%) | | Thickness tolerance (mm) | | Wedge tolerance (microns TIR) | |
|---|---|---|---|---|---|---|---|---|
| | spec | value | spec | value | spec | value | spec | value |
| C1 | 3 | 1 | 0.08 , 0.09 | 0.02 , 0.04 | 0.3 | 0.2 | 100 | 51 |
| C2 | 3 | 1.8 | 0.16 , 0.08 | 0.08 , 0.01 | 0.5 | 0.3 | 100 | 72 |
| ADC1 | 4 | 3.5 | 0.22 , 0.15 | 0.02 , 0.003 | 0.3 | 0.15 | 140 | 1 |
| ADC2 | 4 | 2.3 | 0.14 , 0.10 | 0.001, 0.01 | 0.3 | 0.15 | 140 | 113 |
| C3 | 10 | 1.6 | 0.15 , 0.10 | 0.05 , 0.07 | 0.15 | 0.09 | 100 | 58 |
| C4 | 5 | 3 | 0.11 , 0.19 | 0.01 , 0.03 | 0.15 | 0.07 | 100 | 76 |

| | Overall figure error (waves P-V) | | Surface slope error (rms microradians) | | High-frequency figure error (nm rms) | | Surface roughness (nm Ra) | |
|---|---|---|---|---|---|---|---|---|
| | spec | value | spec | value | spec | value | spec | value |
| C1 | 0.5 | 0.09 | 1.5 | 1.25 , 0.6 | 6 | 3 , 5 | 2 | 0.5, 0.6 |
| C2 | 2 | 1 , 0.2 | 5 , 2 | 6.3, 1.5 | 6 | 4 , 2 | 2 | 1.9, 0.2 |
| ADC1 | 1 | 0.7, 0.6 | 2 | < 2, < 2 | 6 | 3 , 2 | 2 | 0.7, 0.6 |
| ADC2 | 1 | 0.5, 0.6 | 2 | < 2, < 2 | 6 | 4 , 3 | 2 | 0.9, 0.6 |
| C3 | 4 | 1.7, 0.7 | 5 , 2 | 4.97, 1.9 | 6 | 5 , 2 | 2 | 1.1, 0.9 |
| C4 | 2 | 0.3, 0.1 | 3 | 2.2, 2.4 | 6 | 2 , 2 | 2 | 0.7, 0.5 |

*Note: if specs differ for either front or back side of lens, both values are given*

*\* Inhomogeneity may be specified over subapertures, or have low-order power term removed*

The finished diameter of the lenses is not shown in      Table 2, although all final measured diameters met their spec. The Table also does not show the index of refraction of the lens glasses. The actual index of refraction as a function of wavelength, aka melt data, was measured in the glass blanks by the glass manufacturers. Only the C1, C2, ADC1, and ADC2 blanks were measured: analysis showed that the optical design is not sensitive to expected glass index variations in C3 and C4, and so the measurements were not taken. In most cases the index data met the DESI specification for absolute index. The glass used to make ADC1 had an absolute index slightly out of its tolerance, but analysis showed that DESI could accept the glass with an insignificant performance hit.

## 5. OPTIMIZATION OF THE LENS SPACINGS

After the lenses are polished, their as-built lens parameters will differ slightly from their prescribed values. For example the radius of curvature of a particular lens may be slightly larger than what is designed, although still within tolerance. Therefore the performance of the total corrector system will be slightly degraded. However, small errors in some lens parameters – radius of curvature, lens thinkness, and glass index – can be largely compensated by a slight respacing of the lens separations. In fact, after all six lenses are polished, we have knowledge of their as-built lens parameters, and we can therefore reoptimize the spacings of the optical design to recover some performance. Note that generally these are the only parameters that can be compensated; other errors such as surface figure error are therefore not included in the reoptimization exercise.

The Zemax optical model is updated to include the as-built values for all lens radii, thicknesses, and glass index data. The model allows the overall corrector position to vary, as well as the lens positions for C1, C2, C3 and C4. Analysis shows that these are sufficient, and ADC1 and ADC2 and the focal plane can stay fixed. Additionally, the Zemax merit function weakly contrains these lens positions to keep them close to nominal, so that the reoptimization does make large changes for only slight performance gains.

The model's merit function is set up to minimize the average image spot size across the full field of view and bandpass. To do this, we must use a sufficient density of field points across the field of view (709), of wavelengths within the bandpass (6), and of Zenith angles within their range (4). An early optimization attempt failed because we used too few wavelengths, resulting in a design where some wavelengths performed very poorly.

The respacing exercise successfully improves the performance of the design. Table 3 shows the average image spot size during several steps in the process. At first the spot size is strongly degraded by the as-built values. However, a simple refocusing of the entire corrector compensates for most of the increase, thus demonstrating that most lens parameter errors cause a straightforward power error. Then an appropriate respacing recovers the original nominal performance completely. Note that the optimization is run for several cases, each with different amounts of weighting that constrains the lens positions to stay close to nominal. It can be seen here that weighting the merit function can achieve a balance between small spot sizes and small lens shifts away from nominal. (The data in Table 3 is from an early version of the optimization and so does not quite match the data shown later.)

Figure 4, Figure 5, and Figure 6 compare the image spot size performance of the nominal optical model and the respaced model, as a function of field position, wavelength, and Zenith angle, respectively.

Table 3. The average spot size is recovered by respacing the lenses.

| | Cases: | ave rms spot radius (microns) | diff. from nominal | lens shifts from nominal positions (mm) | | | | |
|---|---|---|---|---|---|---|---|---|
| | | | | defocus | C1 | C2 | C3 | C4 |
| *Progression* | nominal model | 15.7 | 0 | 0 | 0 | 0 | 0 | 0 |
| | make all "as-built" changes | 79.7 | 64.0 | 0 | 0 | 0 | 0 | 0 |
| | reoptimize defocus only | 16.6 | 0.9 | -0.411 | 0 | 0 | 0 | 0 |
| | reoptimize all lenses — 0.01 weights | 15.9 | 0.2 | -1.102 | -0.155 | 0.239 | 0.113 | -0.277 |
| | reoptimize all lenses — 0.002 weights | 15.7 | 0.1 | -1.173 | -0.128 | 0.266 | 0.337 | -0.528 |
| | reoptimize all lenses — no weighting | 15.7 | 0.0 | -0.706 | -1.022 | -0.724 | 1.263 | -0.933 |

| Average = 16.29 | Max = 19.53 | Min = 13.11 |
| Average = 15.88 | Max = 19.82 | Min = 12.26 |
| change: -0.41 | 0.29 | -0.86 |

Figure 4. Variation of image spot size across the 3.2 degree field of view. The left image shows the variation of the nominal optical model, and the right image shows the variation of the respaced model. The plot elements are color-coded to make patterns evident: red is high, yellow is low, green is null, outside the designed field of view.

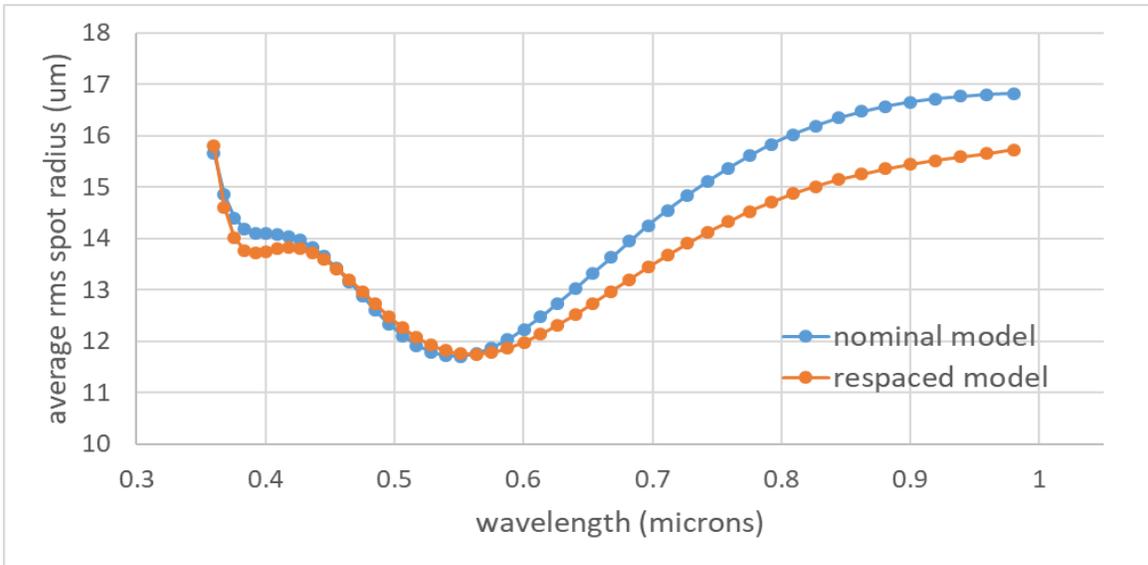

Figure 5.  Variation of image spot size over the DESI bandpass.

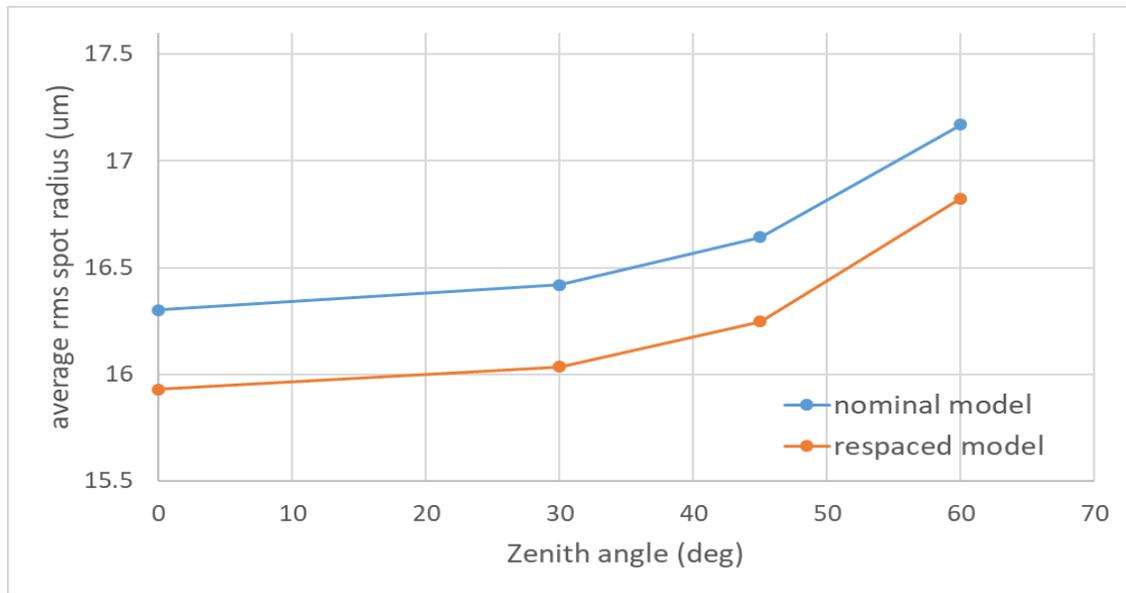

Figure 6.  Variation of image spot size over the range of DESI Zenith angles.

Respacing the lenses results in spot size performance that is as good as the nominal design, on average over the DESI field of view, bandpass, and Zenith angle range.  The respaced model actually performs slightly better, with the average spot size being a few percent smaller.  The reason for the slight improvement is simply due to the fact that the analysis merit function in the model is different from the original optimization merit function that was used to create the design.  The optimization merit function used significantly fewer field points, and they were weighted unevenly to push on trouble spots; thus when the original design is examined with a much denser array of fields, it doesn't perform as well as was first estimated, and reoptimizing with a dense-fields merit function improves it.  The optimization merit function was also structured to constrain any design parameters that are now fixed, e.g. lens thickness, aspheric slope, ADC wedge, etc, which no longer weigh into the analysis.

The respacing optimization focuses on image spot size as its metric, but the DESI corrector optical design also must constrain the angle of incidence at the focal surface to be close to normal. We examine the angles of incidence in the respaced design and find that the maximum change is 0.004 degrees, which is negligible.

Table 4 shows the changed positions of the corrector lenses due to the respacing optimization. The changes are all under 1 mm, small enough to be allowed within the range of adjustment within the barrel.

Table 4. Axial positions of the corrector lenses before and after respacing.

| surface | nominal design vertex position | respaced model vertex position | change |
|---|---|---|---|
| hexapod defocus | 0.000 | **0.588** | *0.588* |
| C1 fr | 2424.231 | **2425.007** | *0.776* |
| C1 bk | 2287.877 | **2288.417** | *0.540* |
| C2 fr | 1812.606 | **1813.200** | *0.594* |
| C2 bk | 1767.606 | **1768.480** | *0.874* |
| ADC1 fr | 1577.606 | **1577.606** | *0.000* |
| ADC1 bk | 1517.606 | **1517.456** | *-0.150* |
| ADC2 fr | 1492.606 | **1492.606** | *0.000* |
| ADC2 bk | 1432.606 | **1432.456** | *-0.150* |
| C3 fr | 1232.609 | **1233.199** | *0.590* |
| C3 bk | 1152.609 | **1153.113** | *0.504* |
| C4 fr | 616.090 | **615.927** | *-0.163* |
| C4 bk | 399.238 | **399.007** | *-0.231* |
| FP | 0.000 | **0.000** | *0.000* |

*values are lens vertex positions in mm; in global coordinates, origin at focal surface vertex*

After the reoptimization is performed in Zemax, knowledge of the updated lens positions can be used to reposition the lens cells in the barrel. The barrel contains steel spacer rings at each barrel-cell interface for exactly this purpose: these spacers are machined to the appropriate thickness to place the lenses exactly where they should be[3]. In addition, the adjustable spacers are also used to compensate for dimensional errors in the fabricated barrel, in the fabricated cells, and lens mounting errors into the cells. Altogether, the spacers allow for ±3 mm of position adjustment, enough to compensate for lens position errors (<1 mm) and others.

The respacing is a worthwhile exercise, but it is complicated and susceptible to errors. Furthermore an error would likely not be noticed until the DESI corrector is completely integrated into the Mayall telescope well into the commissioing phase. Therefore the steps of the respacing effort are carefully checked independently by multiple people in various stages: when the as-built lens data is collected, when the spacings are reoptimized, and when the cell spacer thicknesses are calculated.

## 6. CORRECTOR PERFORMANCE VS. SPECIFICATION

The optical performance of the corrector is expected to be excellent, based on the quality of the completed lenses. Early in the DESI program, initial estimates were made for lens errors, and tolerances were derived and flowed to the vendors; in all respects the lenses meet or surpass those tolerances, thus the total corrector lens imaging quality is expected to surpass early estimates as well. Of course, final performance of the corrector depends on the cell and barrel alignments, currently in progress.

While the DESI project does not place strict requirements on the image spot size at the corrector level, it does track spot size in a higher-level budget, and rolled-up estimates of the corrector have remained consistent with early estimates.

## 7. CONCLUSIONS

The polishing and coating of the six DESI lenses are complete, resulting in excellent lenses that are ready for integration into the larger DESI corrector assembly. The lenses meet all of their specifications, mostly with extra margin. The one exception is a figure error on the difficult C2A asphere, which is analyzed and deemed to be acceptable.

With the conclusion of polishing of the lenses, the corrector optical model is reoptimized using the final measured parameters of the lenses, to improve performance of the corrector by respacing the lenses slightly. These updates are provided to the barrel spacers to be machined during lens intergration.

The six lenses are successfully coated with an antireflection coating. All the various lens contracts have been completed. The lenses are now well along in their integration at University College London.

## 8. ACKNOWLEDGEMENTS

This research is supported by the Director, Office of Science, Office of High Energy Physics of the U.S. Department of Energy under Contract No. DE–AC02–05CH1123, and by the National Energy Research Scientific Computing Center, a DOE Office of Science User Facility under the same contract; additional support for DESI is provided by the U.S. National Science Foundation, Division of Astronomical Sciences under Contract No. AST-0950945 to the National Optical Astronomy Observatory; the Science and Technologies Facilities Council of the United Kingdom; the Gordon and Betty Moore Foundation; the Heising-Simons Foundation; the National Council of Science and Technology of Mexico, and by the DESI Member Institutions. The authors are honored to be permitted to conduct astronomical research on Iolkam Du'ag (Kitt Peak), a mountain with particular significance to the Tohono O'odham Nation.